# Light under a matter field microscope

Graham Hugh Cross

*University of Durham,
Department of Physics, Durham DH1 3LE, UK
g.h.cross@durham.ac.uk*

*Textbooks state that the successful application of Maxwell's Equations in physical optics problems requires light to interact with matter where any inhomogeneities are spaced by less than or equal to the wavelength of light – the 'dense' limit. This untested statement is proved correct in the experiment reported here and a precise dense limit is found. A 'matter field microscope' is used. The dense limit appears exactly when the (approximately fixed) optical wavenumber coincides with the (rapidly increasing) spatial frequency for a discontinuous matter field as it develops from a dilute to a concentrated material system. It is inferred that in the dilute material system examined here plane wave methods are only valid over a well-defined, but restricted, transverse dimension which is introduced here as the 'wavewidth'. This appears to reveal a natural sub-wavelength mode structure for light. This result then allows the application of the uncertainty relation to describe a longitudinal spatial restriction of the light when crossing matter density interfaces. Physical optics using 'spatial' rays and 'spatial' photons can then be described on this model for light.*

## Introduction

Following the discovery of weak 'Anderson' localisation of light in granular material systems[1,2] there has been an incentive to discover what this means for refraction and diffraction. Wave packet scattering does not fit into the normal assumptions required to describe these macroscopic optical phenomena. For example to apply a refractive index to problems in classical optics the matter distribution must first be continuous on the scale of the optical wavelength. This allows one to apply 'physical' diffraction theory based on the idea of the propagation of so-called 'Huygens' secondary wavelets. Related to this, the description for light must allow smoothing of any light/matter interactions so that one can think of average current densities induced by average electric and magnetic fields - the 'mean-field' and 'plane-wave' approaches. Textbook methods[3] can then be applied but at the expense of removing physical insight. The experiment presented here involves precise determination of the nature of optical phase delay accumulating through a dilute discontinuous layer medium. The textbook 'rules' are deliberately violated here and the consequences revealed. In fact a rather specific length scale for the 'dense' matter requirement is established beyond which the rules may once again be safely applied. In the process the Huygens wavelet model[4] now acknowledged as an unsatisfactory starting point for diffraction[5] and refraction[6] is replaced by a form of photon model.

## The matter field microscope

It is interesting and insightful to consider the experiment as if one is putting light itself under a microscope. Light is therefore 'placed' at the surface of an optical slab waveguide by means of the evanescent fields produced by the two lowest order slab waveguide modes of orthogonal polarisation states (transverse electric "TE" and transverse magnetic "TM"). The slab modes are excited by unfocussed end-fire coupling of an expanded diverging (fan angle 4°) $TEM_{00}$ mode from a HeNe laser using a Powell lens. This waveguide is part of a dual slab waveguide interferometer system[7]. The waveguide modes are excited with a free space wavelength, $\lambda_0$ = 632.8 nm, equating to a waveguide mode wavelength, $\lambda_{eff} \approx 420$ nm.

The light field is 'illuminated' with a composite matter field comprising water or aqueous buffer and a random surface bound distribution of monodisperse polystyrene nanospheres. The surface density of the sphere distribution acts as the 'magnification factor'. Unlike a conventional microscope the 'image' is the measured effective refractive index of the medium through which evanescent light travels. At and above a critical matter field density (the 'matter field diffraction limit') a mean-field effective medium refractive index, $n_{eff}^{EMT}$ given by simple constitutive rules can be safely applied. Below this limit the light appears 'out of focus' and no clear details are visible. This is as far as we will take this analogy.

For this two component inhomogeneous medium comprising spheres of relative permittivity, $\varepsilon_{sp}$ embedded in a 'continuous' medium with permittivity $\varepsilon_m$ at (for a layer system[8]) an area density $f_A$, the effective medium constitutive properties are given reliably by the Maxwell-Garnett (M-G) formalism[9].

$$\frac{(\varepsilon_{eff}-\varepsilon_m)}{(\varepsilon_{eff}+2\varepsilon_m)} = f_A \frac{(\varepsilon_{sp}-\varepsilon_m)}{(\varepsilon_{sp}+2\varepsilon_m)} \quad [1]$$

The effective relative permittivity $\varepsilon_{eff}$ gives directly the effective medium refractive index through $n_{eff}^{EMT} = \sqrt{\varepsilon_{eff}}$ provided that the material is non-magnetic and the mean-field conditions apply. For spheres in a layer the Wigner-Seitz method gives the average nanosphere centre-centre separation, $\langle S \rangle$ according to $\langle S \rangle = 2r_{ps}/\sqrt{f_A}$ where $r_{sp}$ is the known sphere radius. Substitution into [1] gives a method of determining the expected average sphere separation from the measured effective medium index. An exhaustive survey of the correct application of this mean-field method shows that it converges with the more refined Mie scattering theory methods when 'long wavelength' conditions prevail (small values for $r_{sp}/\lambda_{eff}$ as in this case) and particularly when the permittivity differences between spheres and continuous medium is low[10,11,12,13]. Perhaps counter-intuitively the methods are predicted to become more applicable as the system becomes more dilute[13].

The experimental method uses a dual slab, dual polarisation evanescent wave, waveguide interferometer (DPI)[14] (see Methods Summary). Phase retardations $\Delta\phi_{TE}$ and $\Delta\phi_{TM}$ in the two upper slab waveguide modes yield the index and thickness for an equivalent continuous surface layer by applying continuum electrodynamics (Maxwell's Equations). Notwithstanding that these deduced index and thickness values may be unphysical for a dilute layer we nevertheless want to seek a convergence condition with the M-G type of theory. The M-G model assumes a constant retardation per deposited sphere in any of the known methods. The deduced layer thickness and the average sphere separation for spheres of nominal diameter 30 nm is shown plotted versus $\Delta\phi_{TE}$ in Figure 1(a). Total disagreement with the M-G prediction occurs until $\langle S \rangle \approx 67$ nm, close to the region where the layer thickness makes sense. This critical average spatial frequency, or critical separation, $\langle \nu \rangle_C = 1/\langle S \rangle_C$, coincides with the following correspondence relationship, $1/k = \lambda_{eff}/2\pi = \hbar/p_M = \langle S \rangle_C$, where $k$ is the classical effective wavenumber and $p_M$ the 'Minkowski' de Broglie wave momentum. Whilst the spatial frequency must be increasing rapidly as the sphere separation decreases, $k$ only increases by a factor of very much less than 0.1 % over the experimental range due to the weak perturbation to the optical evanescent field. Since $k$ is essentially constant at all spatial frequencies and the mean-field method only applies at and above this critical frequency, it is safe to say that the appropriate transverse electromagnetic plane wave is fundamentally constrained to a spatial dimension of around 67 nm.

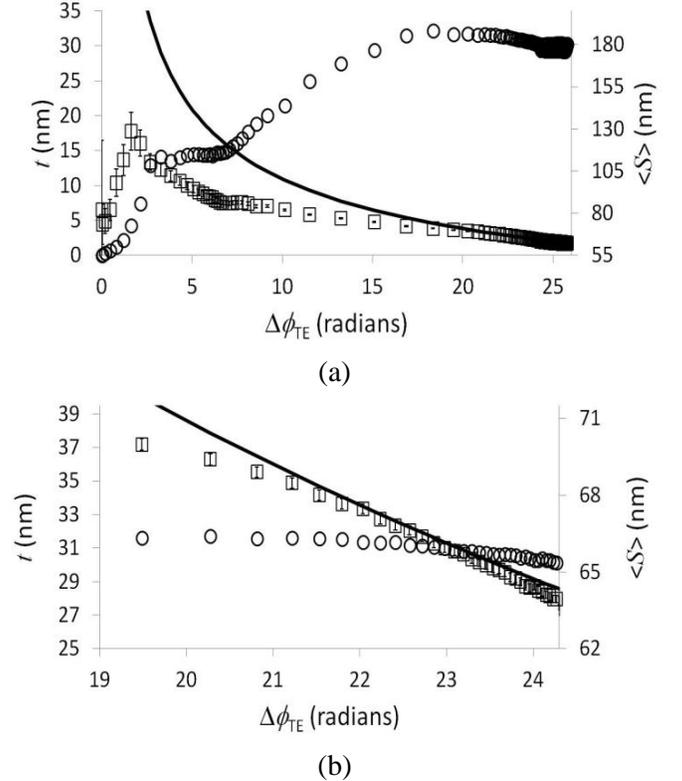

*Figure 1: (a) Deduced layer thickness (circles) and average separation (square) versus TE mode phase retardation for the deposition of 30 nm diameter polystyrene spheres, Maxwell-Garnett (EMT) model average predicted separation shown for comparison (solid line). The region between 3 and 7 radians shows the effect of reducing the pH of the flow medium so that chemi-sorption is encouraged. Higher surface binding rate here seems to be associated with a strong differential decrease in deduced separation whilst the deduced thickness remains constant. (This feature should be considered in the context of the explanations proposed later in the paper where it will be attributable to a large transient increase in layer inhomogeneity.) (b) Detail of the region of agreement with the M-G model. Error bars are calculated using the phase retardation random noise 5σ uncertainty of ±5 mrad and ±3 mrad, for Δϕ<sub>TE</sub> and Δϕ<sub>TM</sub>, respectively.*

A dilute system of 100 nm diameter spheres is now considered to further explore the irrational dilute limit behaviour. The final maximum occupied surface area density is close to 1 % across the waveguide surface with an average sphere separation of ~1,000 nm (Figure 2(a)). This time we avoid using textbook electromagnetism to characterise the layer. We are content to follow the experimental ratio $R = \Delta\phi_{TE}/\Delta\phi_{TM}$ as a function of $\Delta\phi_{TE}$. Fortunately under these experimental conditions there is, in continuum electrodynamics, an approximately linear relationship between $R$ and the deduced effective refractive index of the equivalent surface layer. This provides a powerful tool on which to base hypotheses describing the qualitative behaviour in the dilute region.

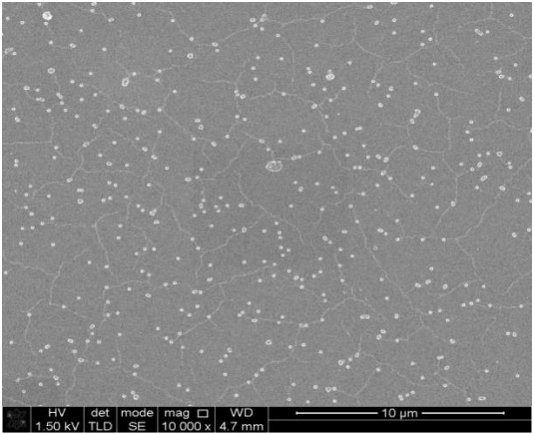

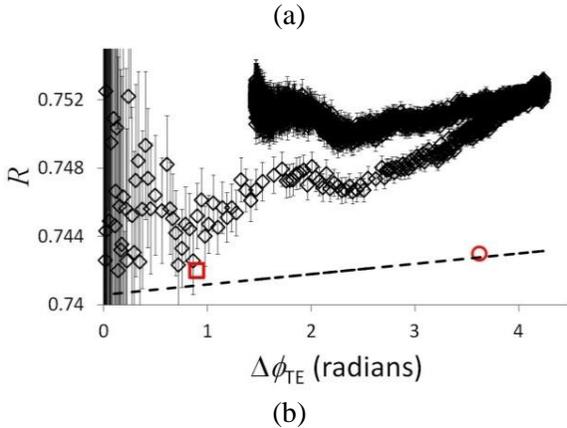

*Figure 2: (a) SEM micrograph of the surface bound 100 nm spheres. (b) R versus TE mode phase retardation (open diamonds for the 100 nm beads shown in (a). Error bars are ±1σ. The simple effective medium model for R versus $\Delta\phi_{TE}$ is shown as the dashed line. The square and circle points are the R values obtained by applying the 100 nm equivalent layer M-G model and the Mie theory model, respectively, using the <u>observed</u> surface coverage (~ 1 % from (a)) as the measure of surface coverage, $f_A$.*

As Figure 2(b) shows, $R$ is either negatively correlated or uncorrelated to rising phase retardation up to around $\Delta\phi_{TE} = 1$ rad. The negative correlation, also seen below 1 radian in the 30 nm sphere data and indeed in all experiments conducted using this material system, is quite difficult to rationalise. Effectively it says that refractive index is <u>decreasing</u> (or sphere separation increasing) as <u>more</u> spheres are deposited. After the deposition peak at $\Delta\phi_{TE} = 4.5$ rad, a washing process removes loosely adhered spheres and the retardation decreases, showing that it is closely related to sphere density. Strikingly however, $R$ remains largely unchanged. We must conclude by inference that 'refractive index' is an indefinable concept at this low coverage. The failure of general extended plane-wave and mean-field methods is once again clearly evident.

**A description for naturally occurring spatially constrained modes of light**

From here the thread of an argument based solely on known physical principles and well known observed phenomena is followed. There are no further distinguishing experimental data to present at this stage. The argument must simply be capable of giving a plausible explanation of both the dilute and dense limit behaviours and of the transition between the two. To accept the validity of the argument requires one only to refer back to the distinguishing experimental results shown above. One must then admit that it is valid to assume that a natural state of light exists described by discrete spatial modes having a transverse constraint, probably given by the rather particular value suggested by the data. The extension of this idea leads one to an understanding of the whole data.

The common feature amongst the models mentioned above is that they assume an unbounded plane-wave model for light. Because the mean-field effective medium theory (EMT) only works well for sphere distributions above some average spatial frequency range, the spatial extent over which an EMT is valid must also be limited.

The development comprises two stages. First, a geometrical ray model for optics is adopted where a finite cross sectional "area of influence" is applied. In the geometrical theory of diffraction (GTD)[15] the ray is simply a direction but here it must be of a certain minimum transverse dimension to explore an effective medium material property. Second, to account for a particle-like diffractive scattering of the ray, a centre of momentum density must be found within it. This demands that the momentum density distribution of the ray be bounded in longitudinal extent. The spatial photon that emerges has a "volume of influence".

**A spatial ray model for propagation**

Light in a broad illumination comprises independent parts, each of which takes an individual route through the medium. Phase retardation accumulates by spatial rays passing through paths with permittivity inhomogeneities ('refractive' paths) where a local effective medium theory may be applied and $R$ is proportional to the local effective permittivity. However, there are also paths resulting from off-axis diffraction that may be free of such inhomogeneities ('geometrical' paths). These also provide phase retardation but the contributory $R$ value is always unity (independent of polarisation).

In the interferometer the measured (forward scattered) phase changes relative to the reference mode field may be given by

$$\Delta\phi_{TE(TM)} = \Delta\phi_r + \Delta\phi_g \qquad [2]$$

where $\Delta\phi_r$ and $\Delta\phi_g$ are the total optical path length refractive and geometrical changes, respectively to the phase of a discrete propagating waveguide mode ray. These comprise incremental phase changes acquired along incremental optical paths, $s$. That part of an incremental path contributing a phase retardation due to a change in the effective refractive index is $\delta\phi_r$ so that $\Delta\phi_r = d \int_0^s \delta\phi_r . dr$ and any part contributing an additional geometrical length $\delta\phi_g$ so that $\Delta\phi_g = c \int_0^s \delta\phi_g . dr$. The coefficients $d$ and $c$ represent the proportions of each type making up the total phase changes.

The incremental refractive phase change component is polarisation dependent and can be expressed as an incremental optical path length change according to

$$\int_0^s \delta\phi_r . dr = k_0 \int_0^s \delta \Delta n_a^T(\vec{r}) l \cdot dr. \qquad [3]$$

Here the propagating ray is treated as having a bounded area of influence, $a$ over which its transverse electromagnetic field can probe any changes in the transverse effective medium refractive index distribution, $\Delta n_a^T$ due to a finite ray-area occupied fraction, $f_a(\vec{r})$ of polystyrene spheres, approximately

$$\Delta n_a^T(\vec{r}) = \Delta\{f_a(\vec{r})(n_{sp} - n_m)\}. \qquad [4]$$

Where $n_{sp}$ and $n_m$ are the refractive indices of spheres and medium, respectively. When $c = 0, d = 1$ giving for the experimental system, $\Delta\phi_{TE}(0) < \Delta\phi_{TM}(0)$ so that their ratio becomes, $R_r < 1$ increasing linearly with ray-area occupied fraction of spheres.

The incremental geometrical phase change component is polarisation independent and can be expressed as an incremental path length change, $\delta l$ according to

$$\int_0^s \delta\phi_g . dr = k_0 \int_0^s \delta l . dr = k_0 \int_0^s (l/\cos\theta) - l . dr. \qquad [5]$$

where $\theta$ is a ray diffractive scattering angle between the original, $l$ and scattered incremental ray paths.

When $c = 1, d = 0$ giving $\Delta\phi_{TE} = \Delta\phi_{TM}$ so that their ratio is always, $R_g = 1$.

The measured value of $R$ is revealed as a weighted average $\bar{R} = dR_r + cR_g$ and if any significant values of $c$ are present, $\bar{R}$ will always be enhanced over the assumed EMT value. If the continuum models are used here, they will always indicate an anomalously high refractive index, as is seen. Dilute systems in particular produce such anomalies where $c \approx d$ indicating that a refractive contribution is as equally likely as a geometrical contribution.

The term, $f_a(\vec{r})$ applies only over the area of influence for the ray. For a transverse ray 'wavewidth', $x$ at least equal to the sphere separation, $(x \geq S)$ the effective ray-area occupied fraction tends towards the mean field value, $(f_a \Rightarrow f_A)$ and the normal EMTs can accurately model the data. As deduced from Figure 1(a), the point at which this occurs is close to $\lambda_{eff}/2\pi = 67$ nm. This suggests that a fundamental lower spatial bound limit for the ray is given by $x \approx S = h/p = 67$ nm. What is contained within this ray area? That is not something to comment on here other than it is required to be a distributed transverse electromagnetic field suitable to apply to Maxwell's equations. However, a two-dimensional centre of momentum density (COM) presumably exists somewhere within the ray bounded area. The usual treatment of the finite spatial photon as a wavefunction comprising a sum of classical plane waves is insupportable on this new model idea. If each photon has to be bounded to such a small region, its direction of propagation would become somewhat uncertain.

**Scattering of spatial photons**

Scattering is now described by a central force theorem between point-like light and matter particles but the ray must now be spatially bound longitudinally. Unless this happens, the transverse centre of momentum density will be distributed uniformly along the ray path and there will be no well-defined light 'point' with which to refer. Scattering would not be describable by any central force mechanism.

A three-dimensional (3D) COM must interact with a local centre of polarisation density (CPD) in the material. The longitudinal (local $z$ coordinate) spatial constraint condition required to specify a 3D COM can be given by the Heisenberg uncertainty relation $\Delta p_M^L \Delta z \geq \frac{\hbar}{2}$ where $p_M^L$ is the longitudinal 'Minkowski' de Broglie wave momentum with $\Delta z$ the COM position uncertainty in the local direction of propagation. It is changes to this component of momentum that are measured in experiments. In the limit $\Delta p_M^L \Rightarrow 0$, $\Delta z \Rightarrow \infty$ and the ray becomes fully non-local in longitudinal extent. This is the fully monochromatic ray but in practice, its longitudinal dimension in space-time is governed by the photon formation length (coherence length)[16]. Photons propagating in a vacuum or in uniform continuous media have presumably a maximum longitudinal dimension of this type. Spatial rays in discontinuous media however are 'squeezed' at boundaries that are the interfaces across which the Minkowski momentum flux (or "Maxwell stress tensor") diverges. In the dilute inhomogeneous medium these boundaries are spaced on microscopic scales and can be rather diffuse, $\Delta z$ becomes finite and small (at the expense of an increase in $\Delta p_M^L$ of course) but a three dimensional COM emerges and the central force theorem can be developed.

The geometrical photon introduced here is a 3-D spatially constrained distribution of wave momentum. The transverse dimension is fixed but the longitudinal dimension is determined locally with an uncertainty, $\Delta z$ at each boundary. Sharp boundaries represent rapid divergence in the stress tensor and more 'point-like' properties for the photon. In fact, it is likely that at the boundaries between any two different continuous refractive media the photon length reduces down to the effective wavelength and $\Delta z = 0$.

For a single CPD and COM scattering event we can assume a central force, $\vec{F}_M$ and define it in terms of its effect on the total (linear and angular) photon momentum as usual by

$$\vec{F}_M(\vec{r}, t) = \frac{d\vec{p}_M}{dt} \quad [6]$$

Normal particle scattering theory describes the total change in photon momentum as

$$\Delta \vec{p}_M = \vec{p}_{M,i} - \vec{p}_{M,f} = \int_0^{\Delta \tau} \vec{F}_M \cos\psi \, dt = 2\vec{p}_{M,i} \sin\left(\frac{\theta}{2}\right)$$
[7]

where the angles, $\psi$ and $\theta$ are denoted on Figure 3 (a) and $\vec{p}_{M,i}$ and $\vec{p}_{M,f}$ are the initial and final Minkowski photon momenta. A single scattering event impact parameter, $b$ is shown but it is better to describe a configurationally averaged impact parameter, $\langle b \rangle$ in this case (Figure 3(b)).

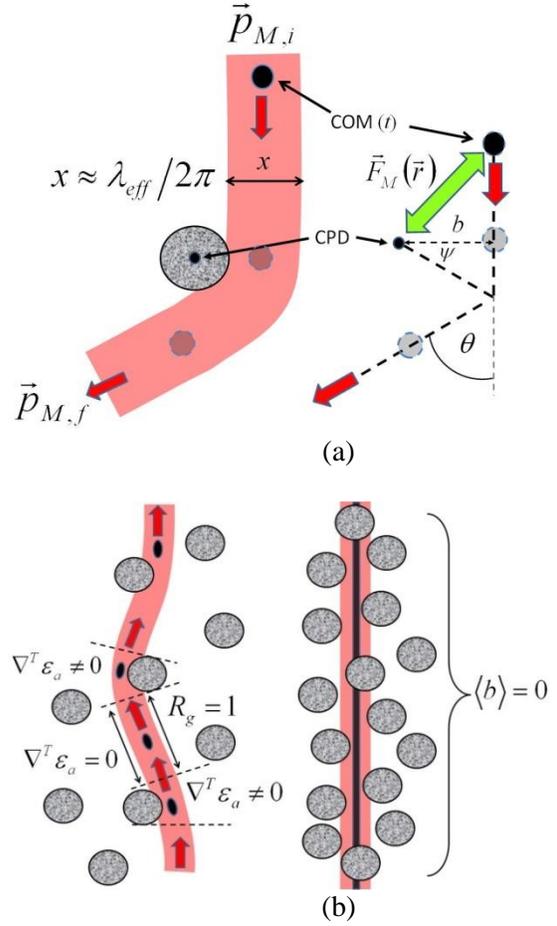

*Figure 3:* (a) A single scattering event with impact parameter, b between a geometrical photon with a propagating centre of momentum density (COM) and a stationary local matter distribution with a centre of polarisation density (CPD). The photon participates in a 'central' force $\vec{F}_M$ as it passes the CPD and is deflected by a scattering angle θ. The photon is a local 'ray' with dimension, x. (*b, left*) Photons with elongated COMs partially undergoing 'zig-zag' diffraction in a sparse CPD distribution. Refractive ($\nabla^T \varepsilon_a \neq 0$) and geometrical ($\nabla^T \varepsilon_a = 0$) paths are identified. The dashed lines represent boundaries across which the Maxwell stress tensor diverges - defining a' local' length for the photon. (*b, right*) Rectilinear photon phase velocity propagation with a distributed COM in a concentrated CPD distribution.

Photon diffraction only occurs when $\langle b \rangle \neq 0$ and when $\langle b \rangle = 0$ all central forces are balanced and the photon is un-diffracted. Clearly this occurs at and above some spatial frequency of the matter discontinuities and all contributions from pure ($R_g = 1$) geometrical phase retardations vanish.

The Minkowski formalism for the photon is the correct one because it describes the propagation of wave momentum in matter. Electromagnetic and matter waves exchange and share momentum and co-propagate.

The force density in an isotropic yet inhomogeneous and non-magnetic, non-conductive, dielectric medium is given by [17]

$$\vec{\mathcal{F}}_M(\vec{r}, t) = -\left[-\left(\vec{\nabla} \cdot \vec{P}(\vec{r})\right)\vec{E} + \frac{\partial \vec{P}}{\partial t} \times \mu_0 \vec{H}\right]$$

[8]

where $\vec{P}(\vec{r})$ is the spatially varying polarisation charge density and $\vec{E}$ and $\vec{H}$ are the incident electric and magnetic field amplitudes. These are expected to be non-uniform in the transverse plane of the ray but in what follows we will focus only on the matter field inhomogeneity to keep the development simple. Equating the polarisation charge density with a linear electric susceptibility we can write $\vec{P}(\vec{r}) = \left(\varepsilon(\vec{r})\vec{E} - \varepsilon_0 \vec{E}\right)$ where $\varepsilon(\vec{r})$ and $\varepsilon_0$ are the spatially varying permittivity and free space permittivity, respectively. Substituting this into [8] and using the vector relationship

$$\nabla \cdot (u\vec{A}) = \vec{A} \cdot \nabla u + u(\nabla \cdot \vec{A})$$

the time-independent part of this force density becomes

$$\begin{aligned}\vec{\mathcal{F}}_M(\vec{r}, \partial/\partial t &= 0) \\ &= -\left\{-\vec{E}\left(\vec{E} \cdot \vec{\nabla}\varepsilon(\vec{r})\right) - \vec{E}\varepsilon(\vec{r})(\vec{\nabla} \cdot \vec{E}) \right. \\ &\quad \left. + \varepsilon_0 \vec{E}(\vec{\nabla} \cdot \vec{E})\right\}\end{aligned}$$

[9]

The first term on the right hand side is an intensity dependent force density component that appears only where the ray's field encompasses inhomogeneities in the permittivity distribution. Since the ray field is fully transverse locally, the gradient in permittivity is of relevance only over the ray's transverse plane, and we introduce the transverse differential vector operator

$$\vec{\nabla}^T = \hat{\imath}\,\partial/\partial x + \hat{\jmath}\,\partial/\partial y$$

If we separately integrate this force density term over an area representative of the ray field we arrive at a net force component whose direction and magnitude is governed by the direction and magnitude of the permittivity gradient.

$$\vec{F}_M = \varepsilon_0 \iint E^2 \vec{\nabla}^T \varepsilon_a \cdot dx dy$$

. [10]

Here, $\varepsilon_a$ is the transverse ray area relative permittivity distribution given by $\varepsilon_a = (n_a^T)^2$. The integral is trivially zero when $\vec{\nabla}^T \varepsilon_a = 0$ of course but also when the permittivity gradient distribution is circularly symmetric. In this latter case, all central forces are balanced.

This central force can in principle be related to a transverse photon field that retains the features required for wave-like phase retardation but has an amplitude that decays asymptotically in the radial direction, $r$. This field might be expressed as $\vec{E}_a(\vec{r}) = \hat{a} A(\vec{r})\exp(ikz)/r^\beta$ where $\hat{a}$ is the field polarisation unit vector, $k$ is the local wavenumber ($k = k_0 n_a^T$), $A(\vec{r})$ is the normalised transverse scalar amplitude. The magnitude of the force would then scale with separation between the COM and CPD according to the following relationship

$$\vec{F}_M(r) \propto \vec{\nabla}^T \varepsilon_a \frac{A^2}{2r^{2\beta}} \cos\psi \propto \vec{\nabla}^T \varepsilon_a \frac{A^2}{2r^{2\beta}} \frac{\langle b \rangle}{r}$$

[11]

Although there is no indication of the spatial structure of the field, it is worth noting that if $\beta$ is unity then the central force law is of an inverse square type.

If the spatially constrained photon explores regions of material where $\vec{\nabla}^T \varepsilon_a = 0$ it will not contribute to any net central force and will continue undeflected. As the photon approaches a region where $\vec{\nabla}^T \varepsilon_a \neq 0$ and when $\langle b \rangle \neq 0$, the net central force develops and the photon is diffracted in an orbital path around the nearest CDP to conserve momentum (the CPD also experiences the Minkowski force and is drawn, correspondingly, towards the photon field). At the critical matter field density when $\langle b \rangle = 0$ the central forces are balanced and photons are undiffracted. The pure geometrical path contributions to the retardation ($R_g = 1$ paths) are now absent and upon further densification the ray area fields extend sufficiently across the local effective medium to report a rational mean-field refractive index; a conventional EMT now applies.

The macroscopic laws of refraction for the photon can therefore be said to have a microscopic diffractive origin and can only be defined and observed in the dense matter limit. Some quantum optics principles are required to complete the description. Reflection is a fundamentally probabilistic phenomenon but total internal reflection and the Goos-Hänchen shift may be deterministic in character according to proposals previously developed[18,19]. In this case, the photon would be bound to a parabolic trajectory at the reflecting interface and the longitudinal shift would be an indication of the magnitude of the central force operating at the interface and the initial angle of approach. In macroscopic diffraction the diffraction angles are determined by an inverse separation force law between photons and the diffracting aperture or edge. A recent review describes the conventional geometrical ray approach to diffraction problems[20].

**Afterword**

Some imaginative postulates have been made for a spatial photon with an electromagnetic spatial soliton as a plausible description[21,22]. It is not the purpose here to favour any of these emerging models, merely to show that there is now a demonstrable reason to restore the propagating quantised 'particle' field in favour of the often cautiously taught secondary spreading wavelet field as the microscopic model of light for diffraction, refraction and reflection. Huygens' general principle however can be retained it seems. Enders shows how a unified Huygens' Principle accommodates sharp, non-spreading, wavefronts[23]. The classical central force proposed here is perhaps a modern embodiment of Newton's original conjecture of a material body force acting on 'Rays of Light'[24].

**Methods Summary**

Aqueous phosphate buffer suspensions of nano-spheres flow over the waveguide surface where the spheres become bound. An active liquid crystal half-wave plate rotates the input light linear polarisation between orthogonal states at a duty cycle of 50 Hz. The evanescent fields interact with the depositing spheres, progressively retarding the waveguide mode phase fronts. The slab waveguide mode fields propagate to the output facet where they now can interfere with the reference waveguide mode fields. In the far-field plane a Fourier transformation of a Young's fringe intensity pattern, imaged by a CMOS camera, yields the average mode field phase retardations for TE and TM polarisations. Each data point is the average over 1 second of individual time correlated image samples. The waveguide total reflection mechanism and detection principle act to filter out and deliver only forward scattered light. The Young's fringe pattern represents interference between sample and reference light field wavefronts each containing spatially incoherent phase contributions from light scattered in the plane of the waveguide mode field[25]. This does not disrupt fringe formation of course which only represents the <u>average</u> phase <u>differences</u> between the two[26].

**Acknowledgements** I thank M. Swann and M. Szablewski for critical reading of the manuscript. The data for the 30 nm spheres was kindly provided by M. Swann of Biolin Scientific, Manchester, UK. The 100 nm sphere data was collected by D. Carter.

# Supplementary information

1. **Experimental procedure**

The dual slab optical waveguide silicon chip interferometer has overall area dimensions of 6 × 24 mm. Layers of deposited silicon oxynitride on the silicon wafer form a stack of transparent dielectrics with the first, second and third depositions forming a reference slab waveguide with the second layer refractive index higher than those to either side. There is no lateral patterning. A fourth high refractive index layer of approximate thickness 1 μm forms the upper, sample slab waveguide. A further low index layer of 2 μm thickness is deposited over the whole substrate and then etched away to expose the surface of the fourth layer in two sample 'wells'. The dimensions of each well is 1 × 15 mm where the sample length, $L$ = 15 mm. Experiments can be conducted separately on each well if required. The intention is to provide a 'two slit' source to form Young's interference fringes from the two waveguide mode light fields emerging from the end face of the chip. The slab waveguides are designed to be single mode in each of the two orthogonal polarisation states ('TE' and 'TM'). Light from a helium neon single transverse mode laser is expanded by a Powell lens to provide a 'letterbox' intensity distribution at the waveguide input plane of dimensions greatly in excess of the waveguide layer area. This produces around -26 dB total power coupling to each slab waveguide but this is of no concern. The advantages are that the coupling is extremely stable against mechanical and thermal instabilities and further, the input polarisation state can be rotated by a simple liquid crystal half wave plate (which may produce refractive spatial aberrations) without disturbing the coupling efficiency.

The spatial distribution of the interference fringes, imaged by a CMOS camera, contains the average phase information, relative to the reference waveguide modes, carried by the sample modes. Fourier transformation of the spatial intensity distribution into spatial phase distribution is then carried out.

The evanescent fields of the TE and TM waveguide modes above the sample waveguide surface decay at different spatial rates according to an exponential law; they have different waveguide mode dispersion curves. Optogeometrical changes above the surface redefine the waveguide modes through interaction with the evanescent fields and this forms the basis of measurement. The intention is to use the mode dispersion difference to resolve the effective thickness and refractive index of equivalent continuous layers introduced to the surface.

For this, one needs to know the mode effective refractive indices of the sample waveguide in the absence of the layer. These can be calculated using standard transfer matrix methods once the sample waveguide optogeometrical parameters are known. The sample waveguide parameters are obtained by a calibration process in the apparatus by exchanging pure water for 80 % (v/v) ethanol in water refractive index over the waveguide surface. Both of these materials have well known refractive indices and there is no thickness to determine because the evanescent fields decay with $1/e$ amplitudes of around 140 nm whereas the calibrant extends to more than 120 μm (includes the thickness of a fluoropolymer gasket) above the surface. Use of the transfer matrix method for a typical calibration analysis yields the following data:

| Layer | Refractive Index | Thickness (μm) |
|---|---|---|
| Sample waveguide core | 1.524 | 0.978 |

Knowing the chip optogeometrical parameters it is simple to use the transfer matrix method to calculate the 'zero position' waveguide mode effective indices, $N_{TE,0}$ and $N_{TM,0}$.

Experiments can now be designed to deposit material onto the waveguide surface and recalculate the waveguide mode effective indices at times, $t$ during the deposition, $N_{TE(TM),t}$ using the phase changes obtained from the interference fringe pattern, $\Delta\phi_{TE(TM)}$ as its intensity distribution changes. These are related to the waveguide mode effective indices (the eigenvalues) through;

$$\Delta\phi_{TE(TM)} = (2\pi/\lambda_0)L(N_{TE(TM),0} - N_{TE(TM),t}). \text{ [S1]}$$

Note that retardations (negative change) and advancements (positive change) of the relative phase are readily seen in this experiment. An equivalent layer formed at the surface is included in the model that finds the infinite combination of equivalent layer thicknesses and refractive indices that matches the new eigenvalue for each polarisation state. The two eigenvalues coincide at a unique pair of equivalent layer thickness and index values.

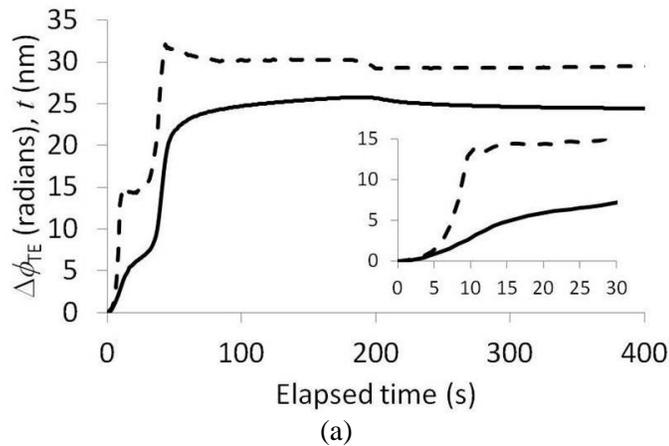

(a)

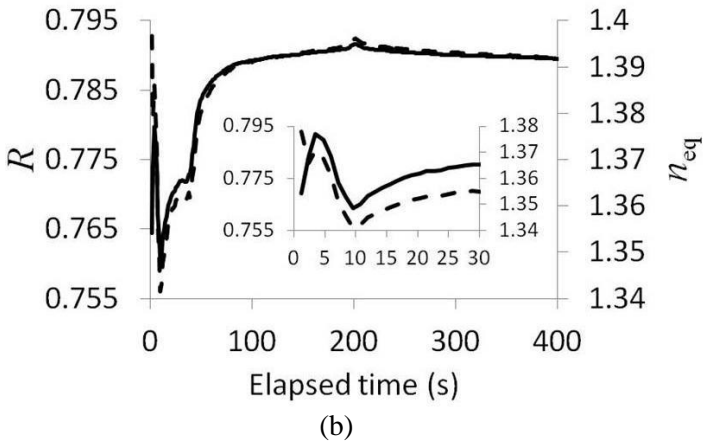

(b)

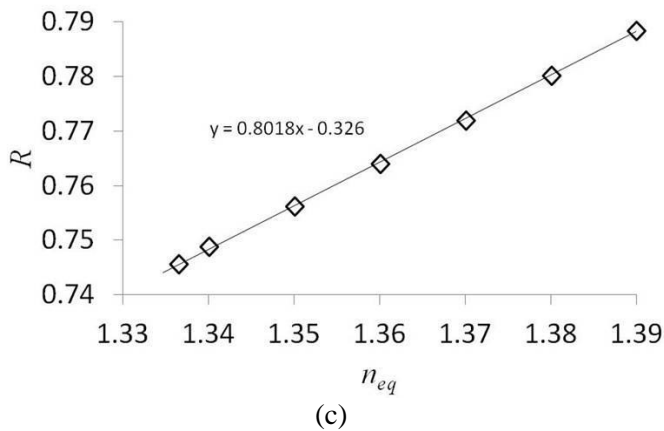

(c)

**Figure S1:** (a) *Deduced equivalent layer thickness(dashed line) and $\Delta\phi_{TE}$ versus deposition time for the deposition of 30 nm spheres*. (b) *Deduced equivalent layer effective refractive index (solid line)and raw data ratio, R versus deposition time for the deposition of 30 nm spheres.* (c) *Calculated R versus equivalent layer effective index for a notional 30 nm equivalent layer.*

Use of the usual Maxwell Equations approach to develop the transfer matrix method for multilayers leads to some important approximate 'rules' that can be helpful in understanding how the technique works. To illustrate this, the data used for Figure 1 in the Article is re-cast here in terms of the raw data collected as a function of time during the experiment. Figure S1(a) and its inset shows how the equivalent layer thickness is deduced as the phase changes accrue. To first order, one can correlate deduced thickness with absolute phase change, regardless of whether this represents the physical situation. This is the more intuitive of the two aspects. Less intuitively, as shown in Figure S1(b) and its inset, the deduced equivalent layer effective refractive index is not directly correlated with phase change alone but almost directly with the phase change ratio, $R = \Delta\phi_{TE}/\Delta\phi_{TM}$. In fact it is possible to calculate an approximately linear relationship between $R$ and refractive index as Figure S1(c) shows, at least over the experimental range and this provides a very powerful qualitative tool when the analysis gives irrational results. A high ratio normally would indicate a high refractive index. When this appears and the phase retardations are still very small, the model has no option but to predict a high index, very thin equivalent layer. The conflict comes when one has to reconcile a high deduced refractive index with a dilute layer at low values of retardation. The idea behind the model proposed in the Article came from rationalising this observation on the basis of discrete light modes geometrically scattered along the waveguide path.

## 2. The experimental ratio $R$ and uncertainty analysis

The importance of the absolute accuracy of the experimental ratio $R$ is such that a full examination of how it arises in an experiment is needed. This discussion describes sources of noise and drift in the recorded phase changes and their treatment.

During an experiment, the fringe image intensity distribution fluctuates, primarily in response to fluctuations in the relative phase retardation between the interferometer reference and sample waveguide modes. Phase fluctuations can arise from transient fluctuations in the optical density in the evanescent field region, perhaps from non-binding impurities in the flow medium and from thermo-optic fluctuations due to irregularities in the flow velocity. Coupled with the intrinsic detector noise, these fluctuations, which average to zero, give rise to a measurement precision of better than ± 1 mrad. The sample stage temperature is controlled to better than ± 1mK at 20 °C but the sample surface will not be kept in register with this absolute value even though it will be controlled to this precision. In addition to these fast fluctuations, there can be some longer term mechanical drift. Mechanical drift would give a polarisation independent drift effect.

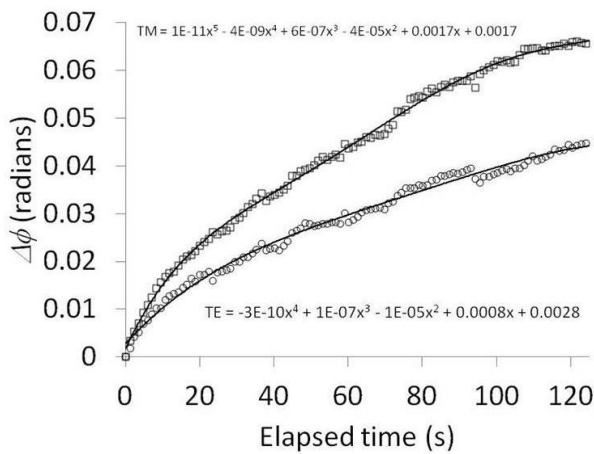

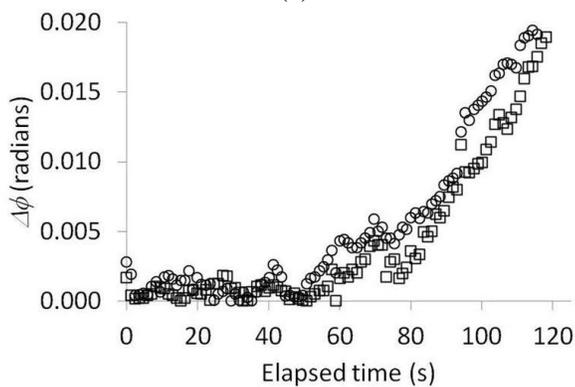

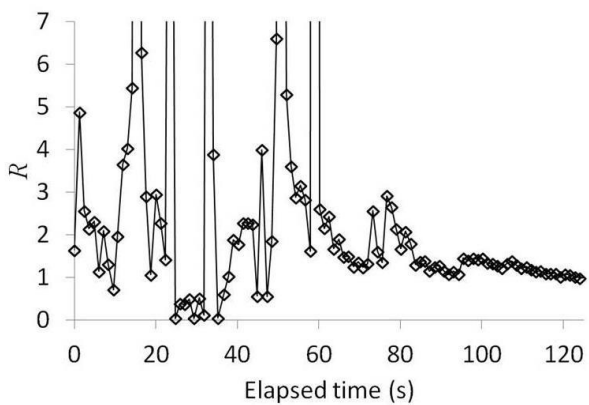

**Figure S2:** (a) *The raw phase retardation data just before the onset of sample introduction (elapsed time shown has been renormalized to show zero start time). The polynomial fits used to find the residuals in this region are shown.* (b) *The residual data shows the emergence of retardations (squares TM, circles, TE) above the noise level after 60 seconds of elapsed time. The noise data for the first 60 seconds of data points is averaged to find a reliable noise standard deviation. The data provides $5\sigma$ noise values as 0.0026 and 0.0047, for TE and TM, respectively.* (c) *Ratio of the residuals at the onset time. Note that the ratio emerges from the noise and tends towards a stable value around unity for the first 20 mrad of retardation.*

Figure S2(a) shows the raw data in a time slot of 125 seconds before any significant phase changes are observed. Some slow polarisation independent drift is observed which could be mechanical in origin. To cancel this out the raw data is fit to high order polynomials from which residuals can be calculated. The residual phase retardation data $\Delta\phi_{TE}$ and $\Delta\phi_{TM}$ at times preceding the observation of an onset of phase changes above the noise is taken and the average of each normalised to zero (Figure S2(b)). The standard deviation $\sigma_{TE(TM)}$ is then found and used as the single point standard error to find the single point error $\alpha_{R_i}$ at later times. This is represented as

$$\alpha_{R_i} = R_i\sqrt{\left[\sigma_{TE}^2/(\Delta\phi_{TE})_i^2\right] + \left[\sigma_{TM}^2/(\Delta\phi_{TM})_i^2\right]}. \quad [S2]$$

The onset of significant data above the residuals indicates that the suspension of spheres has begun to influence the phase retardations. The departure from the residual prediction shows a failure to fit to the third order and higher terms in the polynomials. These have very low coefficients in relation to terms up to quadratic. A combination of bulk suspension and early stage sphere binding will contribute to the results. It is interesting that the ratio emerges from the noise and tends to a value of 1 (Figure S2(c)). No particular significance can be attributed to this at present. Sources of $R = 1$ could be mechanical drift or a high population of geometrical ray paths on the model proposed. This would indicate that sphere binding rather than bulk effects dominates the retardation data and the ratio in the very early stages. More analysis is needed to confirm this effect.

Although a possible drift has been detected in this experiment, the timescale for the complete experiment is around 250 seconds between onset and residual relative phase retardations after sample washing. We note that the initial drift rate is around 30 mrad/min. This is to be compared with the total phase retardations over the experiment time (around 2 minutes) of around 25 radians. Thus, if the drift continued at the original rate (it shows signs of slowing) it would only contribute 60 mrad offset to the final data and could be considered insignificant.

The error bars shown on the average separation data on Figure 1 of the Article are obtained by a propagation of the error taken from that of the experimental ratio as described above. Propagation is a step-wise procedure using the differential method as found in most experimental physics textbooks on measurement uncertainty.

### 3. 21 nm spheres

A number of experiments with spheres of smaller radius have been conducted. The results from one such experiment are presented here (Figure S3). Polystyrene sphere suspensions (in water, 2 mM azide) of 2 % weight/volume (FluoSpheres™ carboxylate-modified, Molecular Probes, Oregon, USA, nominal actual size 21 nm) of an aliquot 200 µl, are diluted with 3.8 ml pure phosphate buffer solution (refractive index = 1.3347, obtained by interferometric calibration against pure water in the same apparatus). The effective medium value of refractive index for the resulting 4 ml suspension is 1.3349. The introduction of this suspension is at a flow rate of 25 µl/min over the surface of the waveguide in a total minimum sample cell volume (governed by the gasket seal dimensions over the chip) of 1.8 µl. The sample introduction time is therefore around 4.5 seconds. During this time a predicted bulk contribution to the phase changes of $\Delta\phi_{TE}(bulk) = 271$ mrad and $\Delta\phi_{TM}(bulk) = 365$ mrad ($R(bulk) = 0.742$) will occur. We may see a transient response corresponding to phase changes originating purely from the bulk medium exchange that are roughly in proportion to this ratio, before the deposition of spheres on the surface completely dominates the observed bulk response.

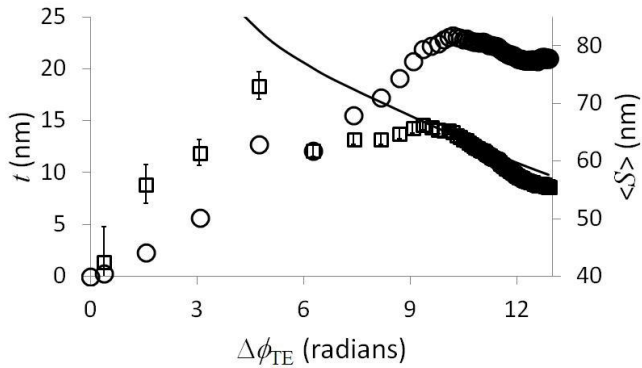

**Figure S3:** *Data from a sphere deposition experiment using (nominally) 21 nm diameter spheres. Deduced refractive index (dashes) converges with the EMT prediction (dash-dot) at $\Delta\phi_{TE} = 10.07$ radians at which point deduced thickness is 23 nm.*

At the onset of agreement with the EMT (at close to 10.1 radians), $<S> \approx 67$ nm and $t = 23.1$ nm. This is the peak thickness and represents the optical maximum layer thickness. Any layer inhomogeneity that re-introduces off-axis scattering must then lower the deduced peak thickness and raise the deduced index (reduce the average separation). The reduction in deduced thickness with an enhancement over the EMT model refractive index after the peak could be evidence of such a layer distribution change, reintroducing $R_g = 1$ type paths and thus bringing back in some enhancement to the apparent index. This effect further emphasises the extrinsic nature of the deduced refractive index as well as the experimental sensitivity to layer morphology.

## 4. Re-examination of the 100 nm sphere data

The proposed scattering model shows that it will be impossible to determine anything specific about the morphology of sphere distributions that are dilute beyond the ray size. Nevertheless, now that a working hypothesis on a bounded ray model is available it should be applied to understand the apparently erratic shape of the $R$ versus $\Delta\phi_{TE}$ curves for the 100 nm diameter spheres. The relevant figure is reproduced here for convenience (Figure S4).

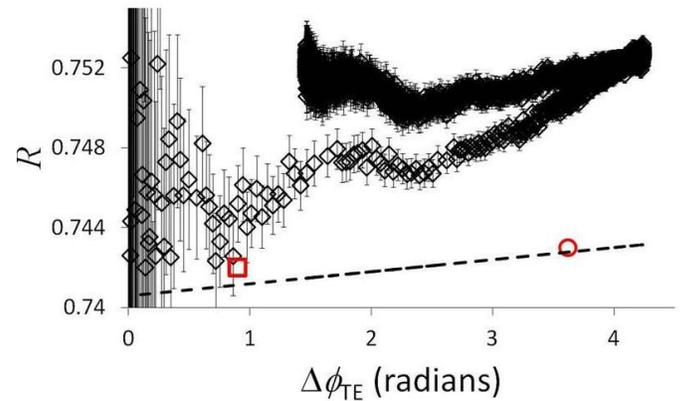

**Figure S4:** *Reproduced from Figure 2(b)*

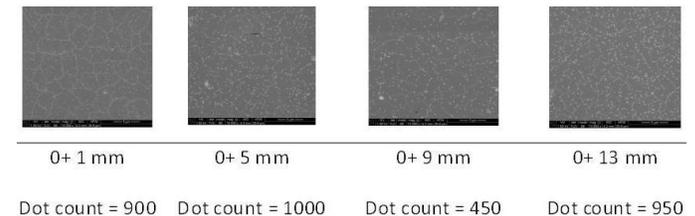

**Figure S5**: *SEM images taken at intervals along the waveguide path following deposition of 100 nm diameter spheres. The overall occupation fraction is constant at ≲ 1%. Image processing software has been used to identify 'dots' on each image to show that relative counts between positions. Most miscounting relates to the SEM legends and text and is thus a systematic error between images.*

The sphere suspension in this case was a dilution in pure water of the 2 % w/v stock, by a factor of 1,000 times. No significant contributions from the bulk effect are expected. These are on the order of a few (less than 10) mrad in each polarisation. The uniformity of deposition at the end of the experiment is shown by the SEM micrographs taken regularly spaced positions along the sample length (Figure S5).

After the usual decline in ratio and a short plateau, the steadily rising $R$ value with phase retardation after around 1 radian now must take into account that the sphere radius is larger than the ray dimension. When the sphere radius is larger than the ray the 'matter microscope' does not have the resolving power to resolve the structure of the light (its spatial frequency will always be too low to resolve the ray dimension). Many rays will, at all scales of dilution, pass fully through single spheres ($b \approx 0$), contributing to refractive paths in some rising proportion, $d = 1 - c$. Paths such as these contribute as if the light has no dimension and as such they might be termed "Huygens paths", $R_H$. We calculate that a Huygens path will contribute $R_H = 0.92$ and independent of sphere occupation fraction. These almost rectilinear paths are almost indistinguishable from deflected $R_g = 1$ paths except for one important feature: they contribute in direct proportion to sphere-occupied area fraction. Deflected paths however can be of any length and scattering angle (representing perhaps the 'mean free path' distribution described in Anderson localisation). Off centre, glancing incidence $R_r$ paths of all scales of local effective medium permittivity will also be present, lowering the average absolute experimental $R$ value. However, adding more spheres will contribute to a more regular, positive correlation as $R_H$ paths become more probable.

The retention of $R$ but not $\Delta\phi_{TE}$ after washing the surface is interesting. If less well bound spheres are being removed one readily understands the loss of $\Delta\phi_{TE}$ as a simple quantitative loss of material. However, the ratio, now that it contains a substantial proportion of Huygens ray paths will remain largely insensitive to phase changes, as predicted by the 'action-at-proximity' Huygens model.

With such an indeterminate distribution of path types it does seem very difficult to justify model fits to the dilute behaviour in any individual experiment. The methods using random walk described in reference 23 in the article may be something for computational physicists to work with. Nevertheless as a starting point and just to illustrate the basic idea an illustrative bi-modal model can be advanced which exchanges just two path types, geometrical and refractive, each contributing the same incremental phase retardation to the average. If one assumes that geometrical paths predominantly contribute to the retardations in the extreme dilute limit and that refractive paths have a very low effective medium contribution in any case the following expression results

$$\bar{R}(\Delta\phi) = \frac{1}{\Delta\phi(max)}\left(\Delta\phi(max) - C\Delta\phi + m(\Delta\phi)^2 + \Delta\phi R_r(0)\right). \quad [S3]$$

Here $\bar{R}(\Delta\phi)$ is the whole path average ratio, $\Delta\phi(max)$ is the expected full layer retardation expected on the EMT models, $\Delta\phi$ is the retardation (either polarisation) accumulating through the experiment, $m$ is the predicted $\bar{R}(\Delta\phi)$ versus $\Delta\phi$ slope value using a simple EMT model and $R_r(0)$ is the limiting ratio at extreme dilution (a small perturbation on the bulk ratio value). The constant, $C$, is a useful scaling factor that carries no physical significance. The model assumes quite crudely that each additional sphere deposited reduces $d$ and increases $c$ monotonically. This elementary function is plotted (Figure S6) to show the idea.

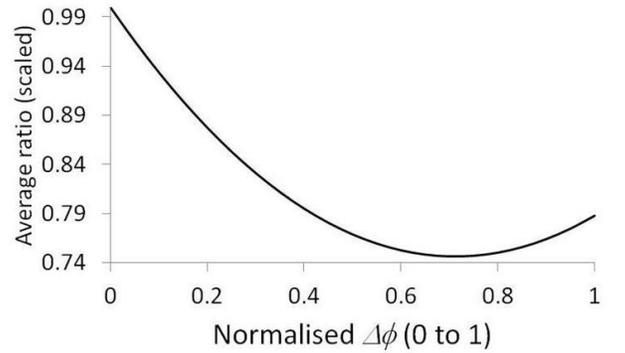

**Figure S6:** *Crude but illustrative bi-modal model for $\bar{R}(\Delta\phi)$ versus normalised phase retardation $d = \Delta\phi = \Delta\phi_{TE(TM)}/\Delta\phi(max)$ showing ideal behaviour when each scatterer increases the refractive path contribution to $\bar{R}(\Delta\phi)$ at the expense of the geometrical path contribution. Parameters for S1: $\Delta\phi(max) = 0.5$, $R_r(0) = 0.744$, $C = 1.1$, $m = 0.25$. The EMT model uses $R_r(\Delta\phi) = m\Delta\phi + R_r(0)$ in the expression $\bar{R}(\Delta\phi) = m\Delta\phi + R_r(0)$ in the expression $\bar{R}(\Delta\phi) = (1-d)R_g + dR_r(\Delta\phi)$ and $R_g = 1$.*

5. **Describing physical optics on this new model for light**

Only in rare circumstances will it be necessary to develop new methods to describe problems in physical optics. The problem presented is one such example. It will be left to interested individuals to verify that the principles given, almost certainly refined and expanded upon, can be rationally applied to <u>all</u> physical optics situations, even though most can be equally well described using the traditional and successful methods. However, by adopting this new description of the underlying origin of physical optics a number of currently contentious issues and paradoxes will be readily resolved.